\documentclass[12pt,a4paper]{article}
\usepackage{amsmath,amssymb}
\usepackage{graphicx}
\usepackage[articletitle=true]{achemso}

\usepackage{xcolor}
\usepackage{paralist}
\usepackage{here}

\usepackage{mathtools} 
\usepackage{booktabs}
\usepackage{siunitx}
\usepackage{bm}

\begin{document}

\begin{center}

\vspace*{1cm}

{\LARGE\bf Corresponding Active Orbital Spaces\\[2ex] along Chemical Reaction Paths}

\vspace{1cm}

{\large Moritz Bensberg and Markus Reiher\footnote{email: markus.reiher@phys.chem.ethz.ch}}\\[2ex]

ETH Z\"urich, Laboratorium f\"ur Physikalische Chemie, Vladimir-Prelog-Weg 2,\\ 8093 Z\"urich, Switzerland\\[4ex]

\vspace{.41cm}

Date: \hspace*{.21cm} 15.02.2023

\end{center}

\vspace{.2cm}

\begin{abstract}
The accuracy of reaction energy profiles calculated with multi-configurational electronic structure methods and corrected by multi-reference perturbation theory depends crucially on consistent active orbital spaces selected along the reaction path. However, it has been challenging to choose molecular orbitals that can be considered corresponding in different molecular structures. Here,
we demonstrate how active orbital spaces can be selected consistently along reaction coordinates in a fully automated way. The approach requires no structure interpolation between reactants and products.
Instead, it emerges from a synergy of the Direct Orbital Selection orbital mapping ansatz combined with our fully automated active space selection algorithm \textsc{autoCAS}. 
We demonstrate our algorithm for the potential energy profile of the homolytic carbon--carbon bond dissociation and rotation around the double bond of 1-pentene in the electronic ground state.
However, our algorithm also applies to electronically excited Born-Oppenheimer surfaces.
\end{abstract}

\newpage
Molecules featuring large conjugated $\pi$-systems, transition metal complexes, or molecules in reactions often feature close-lying, nearly degenerate frontier orbitals. For these cases of static electron correlation, multi-configurational quantum chemical methods provide qualitatively correct electronic
wave functions; examples are the complete active space self-consistent field (CASSCF)\cite{Roos1980, Ruedenberg1982, Knowles1985, Shepard1987},
full configuration interaction quantum Monte Carlo\cite{Booth2009, Booth2010} (FCIQMC) 
and the density matrix renormalization group (DMRG) \cite{White1992, Baiardi2020} approaches. 
Because of the exponential scaling of the number of many-electron basis states with the number of orbitals, these approaches demand a restriction of the size of the orbital basis by defining an active orbital space for a given molecular structure. Then, all possible electronic configurations are constructed for a basis-set expansion of the electronic wave function from these active orbitals. 
As a consequence, the dynamic electron correlation originating from the neglected orbitals must be captured by additional procedures, of which multi-reference perturbation theory\cite{Lindh2020} is a standard choice.

For accurate reaction energies, active spaces must be selected that are consistent for multiple molecular structures along a reaction path, \emph{i.e.}, the active spaces must consist of orbitals that can be considered corresponding between all structures along a reaction coordinate.
Otherwise, the total correlation energy is calculated inconsistently, leading to an erratic behavior of the relative energies. 

Here, we propose a fully automated orbital mapping procedure that ensures consistent orbital spaces between multiple structures along a cut through the Born--Oppenheimer surface. As a byproduct, the mapping procedure identifies the valence orbitals that change significantly. Since orbitals describing bonds that are broken or formed during a reaction cannot be identified in every structure, these orbitals cannot be mapped unambiguously and must be included in the active 
spaces. To ensure consistent active orbital spaces, all of those valence orbitals that are varying along a reaction path are assigned to the active orbital space as soon as one of them is selected for it.

Compared to the active space selection protocol for reactions proposed in Ref.~\citenum{Stein2017}, 
our new approach does not follow orbitals along a reaction coordinate through interpolated structures. Hence, we avoid practical problems that can arise if orbital sets change significantly, preventing direct orbital identification between structures.

For a single molecular structure, active spaces are often selected manually based on expertise or by analyzing the orbital occupation numbers from Hartree--Fock\cite{Pulay1988, Bofill1989, Keller2015a} or M{\o}ller--Plesset perturbation theory\cite{Jensen1988}. A reliable and convenient alternative has been the introduction\cite{Stein2016} of our fully automated active space construction algorithm based on single-orbital entropies\cite{Legeza2003}
obtained from approximate but fast DMRG calculations\cite{Stein2016, Stein2016a, Stein2017, Stein2019, Unsleber2022a}. 
We note that after the proposition of \textsc{autoCAS} in 2016, various other orbital selection approaches have been proposed\cite{Sayfutyarova2017, Bao2018, Khedkar2019}
which, however, are less general, expensive to evaluate, and/or not fully automatically applicable.

Consistent orbital spaces between structures along a reaction coordinate have already been investigated for embedding calculations \cite{Welborn2018, Bensberg2019a}.
For many embedding approaches (\emph{e.g.}, projection-based embedding\cite{Manby2012} or multi-level correlation methods\cite{Sparta2017, Mata2008, Li2010a, Rolik2011, Barnes2019}), an orbital set is selected from the occupied and localized supersystem Hartree--Fock (HF) or Kohn--Sham orbitals. Selecting these orbitals can be automatized through the direct orbital selection (DOS) approach\cite{Bensberg2019a, Bensberg2020a}, which identifies the occupied orbitals that change significantly along a path and selects them for the embedded region. The algorithm provides a bijective map between sets of occupied orbitals that can be leveraged in multi-level coupled cluster calculations\cite{Bensberg2021a, Bensberg2022b}. 

To obtain consistent active orbital spaces for CASSCF, FCIQMC, and DMRG calculations of structures along a path cut through a Born-Oppenheimer hypersurface,
we extend the DOS mapping procedure to include also the virtual valence orbitals. We then combine the automatized orbital mapping with the automatized active space selection algorithm implemented in \textsc{autoCAS}\cite{Stein2019, Moerchen2022}. Finally, we demonstrate that this ansatz leads to fully automatized active orbital selections for chemical reactions in which the multi-configurational character changes 
markedly along a reaction coordinate.

For each structure $L$ along a path, we consider a total orbital set $\{\psi_{iL}\}_\mathrm{tot}$. 
This orbital set is either the set of all occupied orbitals or of all virtual valence orbitals of $L$. We note that the algorithm initially published in Ref.~\citenum{Bensberg2019a} and outlined below is independent of the specific occupation of the orbitals. The orbital maps are independently constructed for occupied and virtual orbitals. The algorithm constructs a map between sets of orbitals rather than between individual orbitals because this allows us to avoid the problem of defining maps between orbitals that change significantly from structure to structure. Furthermore, the mapping criteria are chosen to be invariant under translation and rotation of the molecule since these operations do not change the electronic structure.

The idea of the algorithm is to identify all orbital sets that can be identified unambiguously  
in all structures along the reaction coordinate and then collect all other remaining orbitals in one set $\{\psi_{iL}\}^\text{no-match}$ of non-matchable orbitals.

Each orbital $\psi_{iL} \in \{\psi_{iL}\}_\mathrm{tot}$ is compared to 
all orbitals in all orbital sets $\{\{\psi_{iL}\}_\mathrm{tot}\}$ (each of which is obtained for a structure selected along a path through configuration space) through criteria characterizing its shape 
and spatial localization in the molecule. We explicitly avoid criteria which are directly sensitive to the molecule's structure given in terms of its nuclear coordinates, because we are only interested in the shapes of orbitals. The mapping criteria should only characterize an orbital as a specific bond orbital or lone pair to mirror what a visual inspection of the molecular orbitals in terms of isosurface plots would deliver.

A mapping between two orbitals $\psi_{iL}$ and $\psi_{jK}$ with indices $i$ and $j$ of the structures $L$ and $K$ will be defined if the following condition is met:
\begin{align}
    |t_{iL} - t_{jK}|/\si{E_h} < \tau_\mathrm{kin} \text{ and } \sum_a |q_{iL}^a - q_{jK}^a| < \tau_\mathrm{loc}~.
    \label{eq:mapping_condition}
\end{align}
Here, $\tau_\mathrm{kin}$ and $\tau_\mathrm{loc}$ are predefined thresholds that are chosen to be equal $\tau = \tau_\mathrm{kin} = \tau_\mathrm{loc}$ in practice\cite{Bensberg2019a}.
Furthermore, $t_{\mathrm{iL}}$ is the orbital kinetic energy of orbital $\psi_{iL}$ in Hartree atomic units
\begin{align}
    t_{iL} = \int \psi_{iL}^*(r) \frac{-\nabla^2}{2} \psi_{iL}(r)~\mathrm{d}^3 r~,
\end{align}
and $q_{iL}^a$ is an orbital-wise population assigned to an atom or a minimal-basis-function shell $a$ in the molecule.
The orbital kinetic energy characterizes how compact an orbital is. In principle, the populations $q_{iL}^a$ can be calculated by any population analysis that provides orbital-wise populations, such as the Mulliken population analysis\cite{Mulliken1955}. We employ shell-wise intrinsic atomic orbital (IAO) populations\cite{Knizia2013} because they are basis-set insensitive and distinguish between contributions from different shells of IAOs (\emph{e.g.}, $1s$, $2s$, $2p$). Therefore, shell-wise IAO charges characterize on which atoms an orbital is localized as well 
as the orbital's shape.
The shell-wise IAO populations are given through the projection on the IAOs $\rho_{mL}^a$ (\emph{e.g.}, $p_x$, $p_y$, or $p_z$ for a $p$-type IAO) of the shell $a$, with angular momentum $l_a$, orientation $m$, and structure $L$\cite{Knizia2013},
\begin{align}
    q_{iL}^a = \sum_{ m = -l_a}^{l_a} \left\langle \psi_{iL} | \rho_{mL}^a \right\rangle\left\langle \rho_{mL}^a | \psi_{iL} \right\rangle~.
    \label{eq:shell-wise-iao-charges}
\end{align}

Evaluating the condition in Eq.~(\ref{eq:mapping_condition}) for every orbital pair $\psi_{iL}\in \{\psi_{iL}\}_\mathrm{tot}$ and $\psi_{jK} \in \{\psi_{jK}\}_\mathrm{tot}$ provides an assignment of a set of orbitals $\{\psi_{jK}\}$ to each orbital $\psi_{iL}$. We collect these assignments in a map $s_{LK}$,
\begin{align}
    s_{LK}:~~\psi_{iL} \rightarrow \{\psi_{jK}\}~.
\end{align}
Based on this map, we identify the orbitals that change strongly along the reaction coordinate, \emph{i.e.}, between any pair of structures $LK$.

We split the total orbital set into an initially mappable orbital set $\{\psi_{iL}\}^M_\mathrm{init}$ with $s_{LK}(\psi_{iL})\neq \{\}~\forall K$ (\emph{i.e.}, the orbital is mappable for all $K$) and an orbital set $\{\psi_{iL}\}^\text{no-match}_\mathrm{init} = \{\psi_{iL}\}_\mathrm{tot} \setminus \{\psi_{iL}\}^M_\mathrm{init}$ for which no orbital match can be identified in at least one structure $K$. 
 
\begin{figure}
    \centering
    \includegraphics[width=0.5\textwidth]{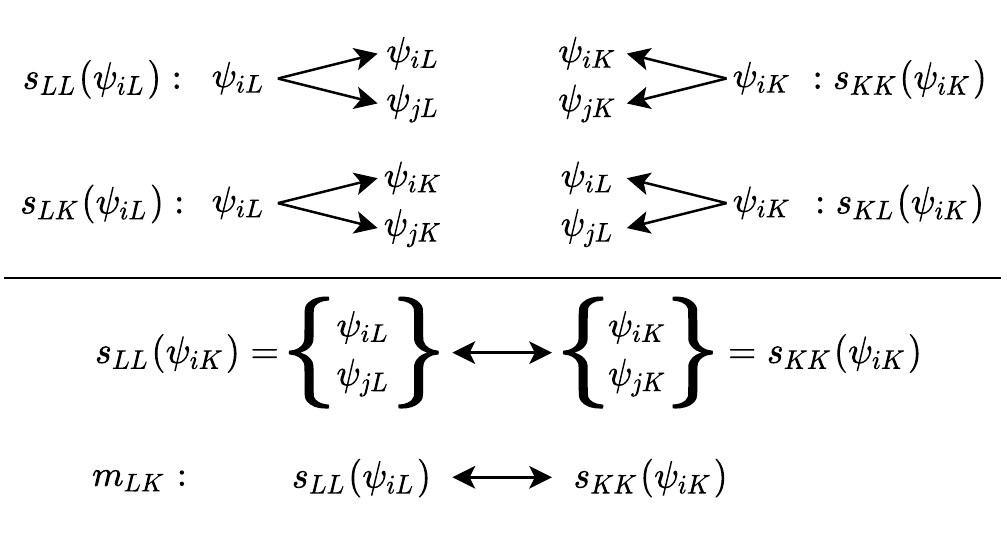}
    \caption{Illustration of the orbital mapping procedure. Each of the orbitals $\psi_{iL}$ and $\psi_{iK}$ is mapped to sets of orbitals through the maps $s_{LK}$ and $s_{KL}$. If these sets are the orbital sets from the self maps $s_{LL}$ and $s_{KK}$, a mapping between $s_{LL}$ and $s_{KK}$ will be defined in $m_{KL}$; \emph{i.e.}, these orbitals can be considered unchanged in structures $K$ and $L$.}
    \label{fig:orbital-set-mapping}
\end{figure}

We must now ensure that the orbital mapping is consistent along the reaction coordinate.
This procedure is illustrated in Fig.~\ref{fig:orbital-set-mapping}. We map the orbital sets obtained from the self-map $s_{LL}(\psi_{iL}) = S_{iL}$ ($\psi_{iL}\in \{\psi_{iL}\}^M_\mathrm{init}$) between structures with the map
\begin{align}
    m_{LK}:~~S_{iL} \rightarrow S_{iK}~.
\end{align}
The condition for assigning a map from $S_{iL}$ to $S_{iK}$ 
in
$m_{LK}$ is that every orbital in $S_{iL}$ is mapped to all orbitals in $S_{iK}$, and not to any other orbital of $K$ through the map $s_{LK}$. Furthermore, the reverse must be true for $S_{iK} \rightarrow S_{iL}$. We then collect all orbital sets that fail this mapping condition ($m_{LK}(S_{iL}) = \{\}$) for some structure combination because they contain an inconsistent orbital mapping. The orbitals in these sets are assigned to the second set of non-matchable orbitals $\{\psi_{iL}\}^\text{no-match}_\mathrm{con}$.

All remaining orbital sets are unambiguously identifiable in all structures according to the mapping conditions. They are mapped through the final bijective map
\begin{align}
    M_{LK}:~~S_{iL} \leftrightarrow S_{iK}~.
\end{align}
The map $M_{LK}$ is bijective and maps orbital sets of structures $L$ and $K$ that are of the same size.
Furthermore, we define a map $A$ between the sets of non-matchable orbitals $\{\psi_{iL}\}^\text{no-match} = \{\psi_{iL}\}^\text{no-match}_\mathrm{init} \cup \{\psi_{iL}\}^\text{no-match}_\mathrm{con}$
\begin{align}
    A_{LK}:~~\{\psi_{iL}\}^\text{no-match} \leftrightarrow \{\psi_{iK}\}^\text{no-match}~.
\end{align}
 Because the total numbers of orbitals in each set $\{\psi_{iL}\}_\mathrm{tot}$ are always the same, 
the sets $\{\psi_{iL}\}^\text{no-match}$ will always have the same size. They contain the orbitals that change during a reaction, \emph{e.g.}, orbitals involved in bond-breaking/bond-formation processes. The maps $M_{LK}$ and $A_{LK}$ combined provide bijective maps between orbital sets containing every orbital of $\{\psi_{iL}\}_\mathrm{tot}$ of every structure.

The mapping algorithm requires orbitals that are transferable between structures. Otherwise, the sets of non-matchable orbitals $\{\psi_{iL}\}^\text{no-match}$ become large, and the mapping is useless. To maximize the transferability of the orbitals, they are localized before the mapping. We found that the intrinsic bond orbital scheme\cite{Knizia2013, Senjean2021} provides highly transferable orbitals for this purpose.

We first localize only the orbitals of one structure and then align all other orbital sets to the already localized orbitals by minimizing the difference in orbital populations
to these template orbitals\cite{Bensberg2020}. Then, all other orbital sets are localized. This increases the chance that all orbital localization procedures converge to similar orbitals. Note that the specific choice of the template orbital set to which all other orbital sets are aligned will influence the final result of the orbital localization. However, it was shown previously\cite{Bensberg2020} that the choice in template orbitals is unlikely to have an effect on the transferability of the orbitals between structures which is the key quantity for the orbital mapping.

The only input parameter for the mapping procedure is the orbital similarity threshold $\tau$ [see Eq.~(\ref{eq:mapping_condition}), and $\tau = \tau_\mathrm{kin} = \tau_\mathrm{loc}$]. However, if we are only interested in a qualitative mapping of the orbitals along the reaction coordinate, we can eliminate $\tau$ 
by requiring that the set $\{\psi_{iL}\}^\text{no-match}$ becomes as small as possible, while it remains reasonable 
for comparing orbital populations 
(in this work $\tau \leq 0.5$)
\begin{align}
    \tau_\mathrm{min} = \min_{\tau} |\{\psi_{iL}\}^\text{no-match}|~.
    \label{eq:tau_min}
\end{align}
Note that $|\{\psi_{iL}\}^\text{no-match}|$ does not necessarily decrease with increasing $\tau$ 
because an increasingly loose mapping through Eq.~(\ref{eq:mapping_condition}) leads to increasingly inconsistent orbital maps $m_{LK}$.

We now combine the mapping procedure with the automated active space selection procedure implemented in \textsc{autoCAS}\cite{Stein2019, Moerchen2022}. \textsc{autoCAS} selects active orbital spaces based on single-orbital entropies calculated by low-cost DMRG configuration interaction (CI) calculations for the entire valence orbital space. These DMRG-CI calculations are carried out by our DMRG program \textsc{QCMaquis}\cite{Keller2015}.

To fully automatize the approach, we interfaced the quantum chemistry software \textsc{Serenity}\cite{Serenity2018, Niemeyer2022, SerenityGitHub} to \textsc{autoCAS}. \textsc{Serenity} implements the orbital-mapping, orbital-alignment, and orbital-localization procedures discussed above.

The complete approach for a set of structures along a reaction coordinate consists of the following steps:\\
(1) Calculate the HF orbitals for all structures.\\
(2) Localize and align the occupied and virtual valence orbitals.\\
(3) Construct the orbital set maps $A_{LK}$ and $M_{LK}$.\\
(4) Select active orbital sets with \textsc{autoCAS} for all structures (or any subset of structures if it is known that more structures will add no additional orbitals to the active orbital set).\\
(5) Construct the active orbital sets for each structure as the union of all sets, which are mapped to a set containing at least one orbital selected by \textsc{autoCAS}.\\
(6) Converge the final active space calculations.

\begin{figure}
    \centering
    \includegraphics[width=0.7\textwidth]{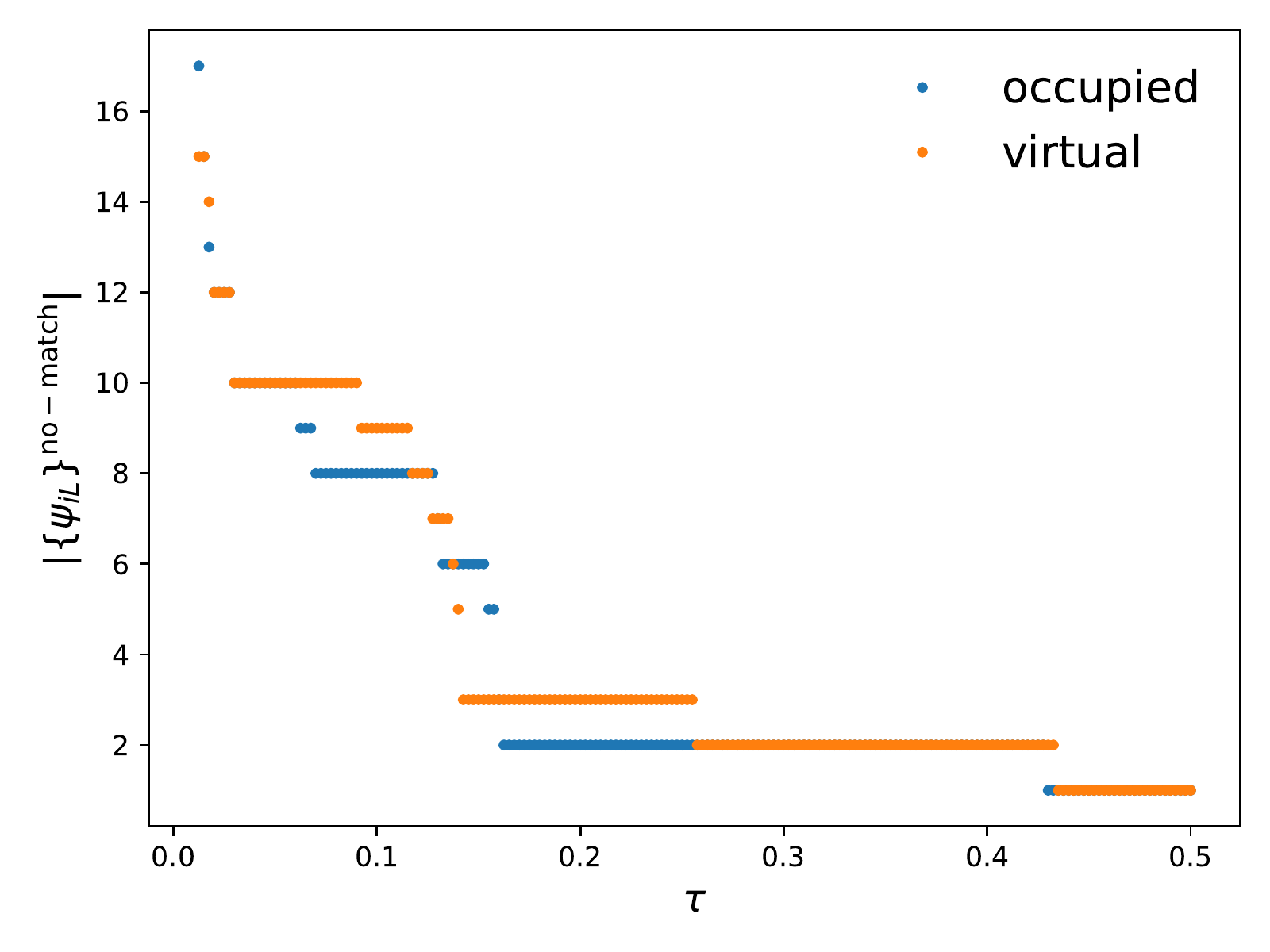}
    \caption{Size of the orbital sets $\{\psi_{iL}\}^\text{no-match}$ for occupied and virtual orbitals with respect to the selection threshold $\tau$.}
    \label{fig:size_vs_tau}
\end{figure}

\begin{figure}
    \centering
    \includegraphics[width=0.8\textwidth]{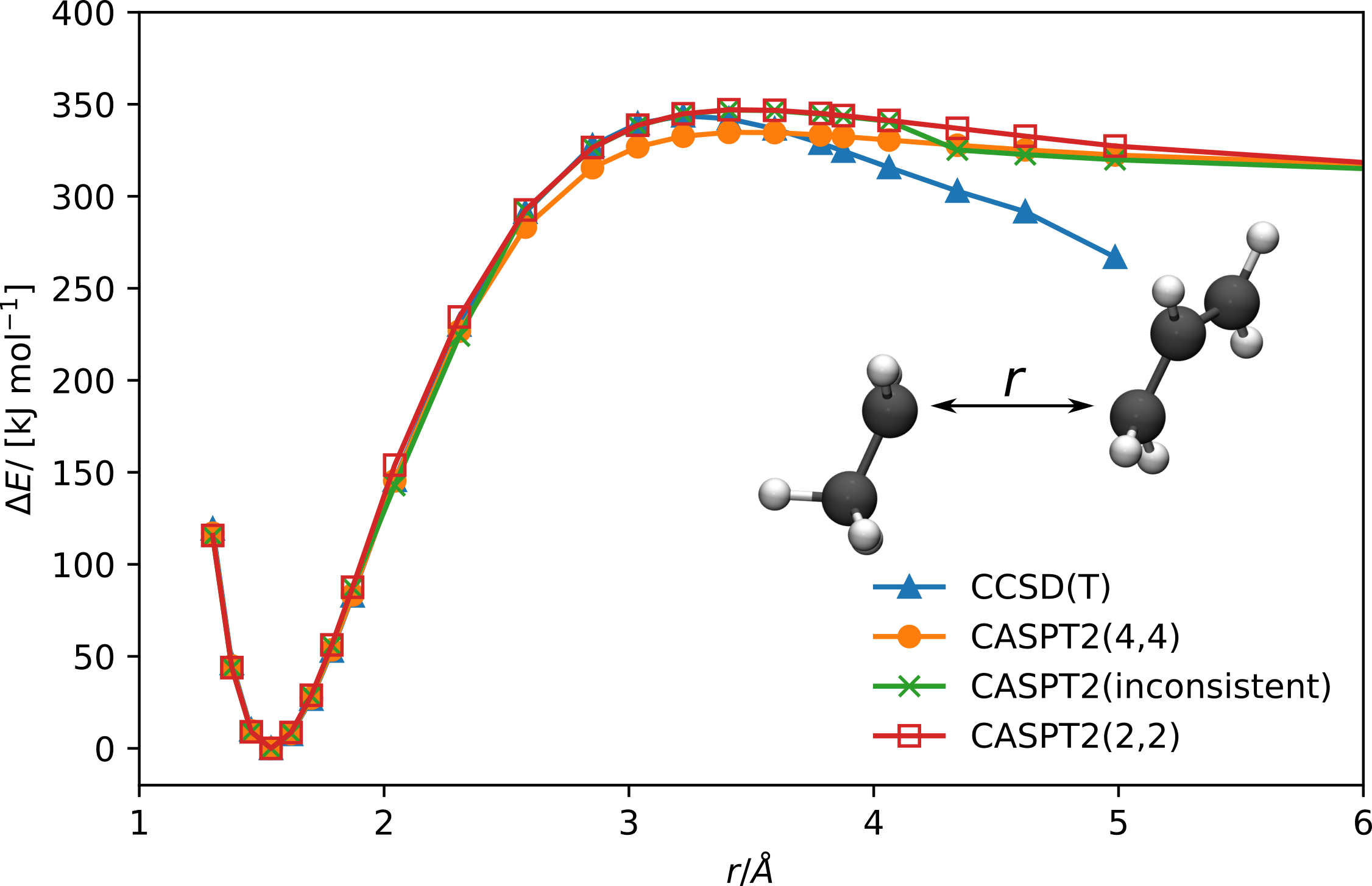}
    \caption{Potential energy profile for the homolytic carbon--carbon bond dissociation in 1-pentene. The energy of the minimum structure is chosen as the zero energy reference point.
    For CASPT2(4,4), the active space was selected for the dissociated fragments and then transferred with the orbital mapping to all structures along the reaction coordinate. For CASPT2(inconsistent), the active space was selected with \textsc{autoCAS} for each structure along the reaction coordinated without ensuring its consistency. For CASPT2(2,2), only the HOMO and the LUMO are included in the active space.
    }
    \label{fig:potential_energy_plot}
\end{figure}

To demonstrate that active orbital spaces can quickly become inconsistent if the electronic character changes significantly along a reaction coordinate, we study the case of homolytic bond dissociation. For stable intermediates such as reactants, a single determinant often describes the electronic structure well. However, a multi-configurational description of the wave function is required upon dissociation. As an example, we investigate the potential energy curve of the homolytic bond dissociation of 1-pentene, as shown in Fig.~\ref{fig:potential_energy_plot}. The structures of the potential energy curve are obtained through curve optimization of the minimum energy path between the bonded and fully dissociated 
fragments\cite{Vaucher2018}. This was carried out with the program \textsc{Readuct}\cite{Brunken2022} applying the PBE exchange--correlation functional\cite{Perdew96} with spin-unrestricted orbital optimization, Grimme's D3 dispersion correction\cite{Grimme2010a}, Becke--Johnson damping\cite{Grimme2011}, and the cc-pVDZ basis set\cite{Dunning1989} in \textsc{Turbomole}\cite{turbomole741} raw data calculations.
We selected 22 structures along this potential energy curve, which are provided in the Supporting Information. For these structures, we
calculated the localized and aligned spin-restricted HF orbitals in the cc-pVDZ basis set\cite{Dunning1989}.  We selected the active orbital space with \textsc{autoCAS} only for the separated product to reduce  the number of DMRG-CI calculations. The separated product showed the strongest multi-reference character, and its active orbital space already contains all orbitals that are needed in the active orbital space for all structures along the reaction coordinate.
Furthermore, we chose a low bond dimension of 250 and a low number of (back-and-forth) sweeps of only 5 for the DMRG-CI calculations, which limit the computational overhead of the active space selection but provide a qualitatively correct description of the single-orbital entropies (illustrated in Tab.~\ref{tab:orbital_entropies} for the highest eight single orbital entropies). The single-orbital entropies are converged up to the fourth digit compared to DMRG-CI calculations with increased bond dimension (up to 1200) and increased number of DMRG sweeps (up to 10).
We then transferred the active orbital space to all other structures.
The selection thresholds for occupied and virtual orbitals were optimized according to Eq.~\ref{eq:tau_min} (occupied: $\tau_\mathrm{min}=0.5$, virtual: $\tau_\mathrm{min}=0.5$). To demonstrate that the orbital mapping procedure is insensitive to the explicit choice of $\tau$, we provide a plot of $|\{\psi_{iL}\}^\text{no-match}|$ with respect to $\tau$ in Fig.~\ref{fig:size_vs_tau}. The size of $\{\psi_{iL}\}^\text{no-match}$ is stable at 1
for the virtual and occupied orbitals before it increases for values of small $\tau < 0.435$.

\begin{table}[]
    \centering
    \caption{The eight highest single-orbital entropies calculated with DMRG-CI for increasing bond dimension ($m$) and number of DMRG sweeps ($w$) for the product of the homolytic carbon--carbon bond dissociation of 1-pentene (both parameters are denoted as $m/w$ in the 2nd to 5th column).}
    \begin{tabular}{l|r r r r}
    \toprule\toprule
  orbital index & 250/5 & 800/5 & 800/10 & 1200/5\\ \midrule  
  5  & 1.06029 & 1.06025 & 1.06025 & 1.06025 \\
  19 & 1.01578 & 1.01574 & 1.01574 & 1.01574 \\
  23 & 0.99425 & 0.99435 & 0.99435 & 0.99435 \\
  32 & 0.33259 & 0.33270 & 0.33270 & 0.33271 \\
  22 & 0.20515 & 0.20517 & 0.20517 & 0.20517 \\
  34 & 0.11351 & 0.11355 & 0.11355 & 0.11355 \\
  28 & 0.11333 & 0.11337 & 0.11337 & 0.11338 \\
  21 & 0.10647 & 0.10652 & 0.10652 & 0.10652 \\
  \bottomrule\bottomrule
    \end{tabular}
    \label{tab:orbital_entropies}
\end{table}

\begin{figure}
    \centering
    \includegraphics[width=\textwidth]{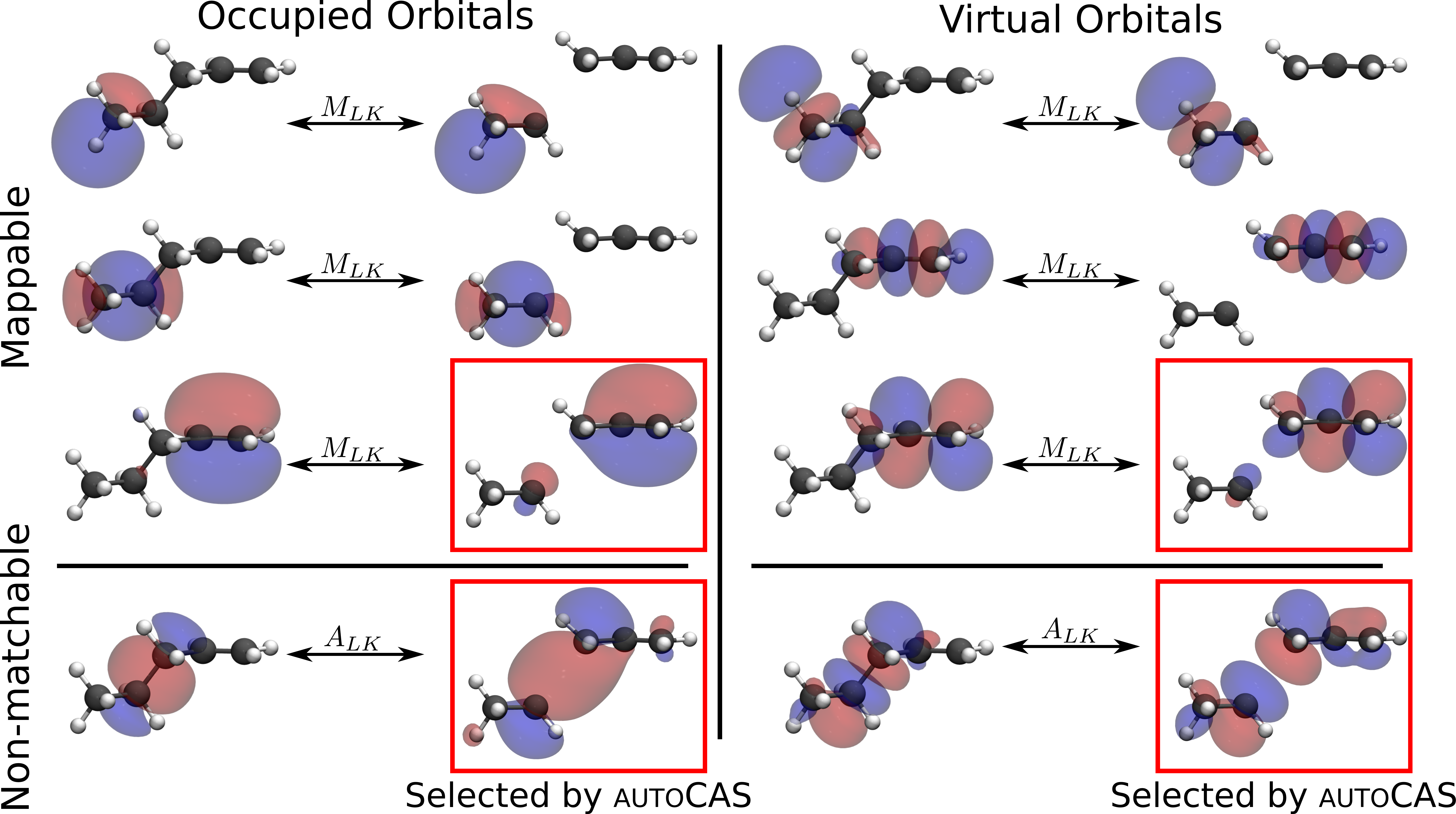}
    \caption{Orbital set mapping through $M_{LK}$ and $A_{LK}$ for valence virtual and occupied orbitals of 1-pentene in its equilibrium and almost dissociated structures. Orbital isosurfaces are shown for a value of $\pm 0.025~$a.u. The red boxes highlight the orbitals selected by \textsc{autoCAS}. The orbitals for the 4th (equilibrium) and 12th structures are shown (the structure indices increase with increasing carbon--carbon internuclear distance). All structures are provided in the Supporting Information.
    }
    \label{fig:orbitals}
\end{figure}

The orbital mapping of the localized virtual and occupied valence orbitals is shown in Fig.~\ref{fig:orbitals} for a selection of orbitals. In this example, every orbital set from the maps $A_{LK}$ and $M_{LK}$ has only one element, \emph{i.e.}, the maps provide a bijection between orbitals. The only two non-matchable orbitals are the bonding and anti-bonding carbon--carbon $\sigma$-bond orbitals of the bond broken upon reaction. The localized orbitals are very compact and represent the expected nodal structure of bonding and anti-bonding $\sigma$- and $\pi$-orbitals.

The orbitals selected by \textsc{autoCAS} are highlighted in Fig.~\ref{fig:orbitals} by red boxes. \textsc{autoCAS} identified the bonding and anti-bonding $\sigma$-carbon--carbon bond orbitals and the bonding and anti-bonding $\pi$-orbitals as statically correlated. These orbitals are then included in the active orbital space for a CASSCF calculation with second-order perturbation theory\cite{Andersson1990, Andersson1992} (CASPT2,
without any IPEA shift\cite{Ghigo2004}). Note that for active spaces with more than about 20 orbitals, which are beyond the capabilities of traditional CASSCF, DMRG-SCF plus subsequent perturbation theory can be used in 
\textsc{QCMaquis}\cite{Freitag2017,Ma2017}.
The potential energy curve calculated with this consistent active orbital space and CASPT2 (without IPEA shift) is denoted as CASPT2(4,4) and shown in Fig.~\ref{fig:potential_energy_plot}.
Furthermore, we show the potential energy curves calculated with CASPT2 and other choices in the active orbital space. We selected the active orbital space with \textsc{autoCAS} from the canonical HF orbitals without ensuring that these spaces are consistent between structures. 
We denote the CASPT2 result calculated with this active space choice as CASPT2(inconsistent). In addition, we selected active orbital spaces as only the highest occupied (HOMO) and lowest unoccupied (LUMO) canonical HF orbital, leading to a CAS(2,2) [denoted as CASPT2(2,2)].
In addition, we show below
the potential energy profiles calculated with coupled cluster theory with singles, and doubles excitations and perturbatively treated triples excitations [CCSD(T)]\cite{Raghavachari1989} as implemented in \textsc{Turbomole}\cite{turbomole741}. The CCSD(T) energies were calculated with spin-restricted orbitals.

The CASPT2 potential energy curves agree with the CCSD(T) potential energy curve for short internuclear carbon--carbon distances ($r < 3.4~\si{\angstrom}$), independent of the choice in the active space. 
For large internuclear distances ($r > 3.4~\si{\angstrom}$), the CASPT2 curves and the CCSD(T) curve show significant differences. All CASPT2 curves converge to a similar, and constant relative energy, which are $314.8~\si{kJ.mol^{-1}}$ for CASPT2(4,4), $312.3~\si{kJ.mol^{-1}}$ CASPT2(inconsistent), and $313.3~\si{kJ.mol^{-1}}$ for CASPT2(2,2). By contrast, the relative energy for CCSD(T) nonphysical decrease. In fact, the CCSD(T) calculation failed for the fragments with the largest carbon--carbon internuclear distance.

The CASPT2 curves show significant differences for internuclear distances between $2.5~\si{\angstrom}$ and $5.0~\si{\angstrom}$. For these distances, CASPT2(2,2) and CASPT2(inconsistent) give very close results with slightly higher relative energies than CASPT2(4,4). The relative energies for CASPT2(2,2) and CASPT2(inconsistent) are higher because of the smaller active space. The active spaces for the structures between $3.7~\si{\angstrom}$ and $4.1~\si{\angstrom}$ contain in the cases of CASPT2(2,2) and CASPT2(inconsistent) only the HOMO and the LUMO. However, for distances $r > 4.3~\si{\angstrom}$, the bonding and anti-bonding $\pi$-orbitals are selected for the active space by \textsc{autoCAS} in addition to the HOMO and the LUMO. Therefore, the CASPT2(inconsistent) potential energy curve shows a sharp kink corresponding to this transition from a CAS(2,2) to a CAS(4,4). The active orbital spaces for CASPT2(inconsistent) show this inconsistency because \textsc{autoCAS} only ensures that the active space is suitable for the given structure and does not take any other structures on the potential energy curve into account.
By contrast, the CASPT2(4,4) potential energy curve is smooth for all internuclear distances investigated here because our algorithm ensures that the active orbital space is consistent between all structures. 

\begin{figure}
    \centering
    \includegraphics[width=\textwidth]{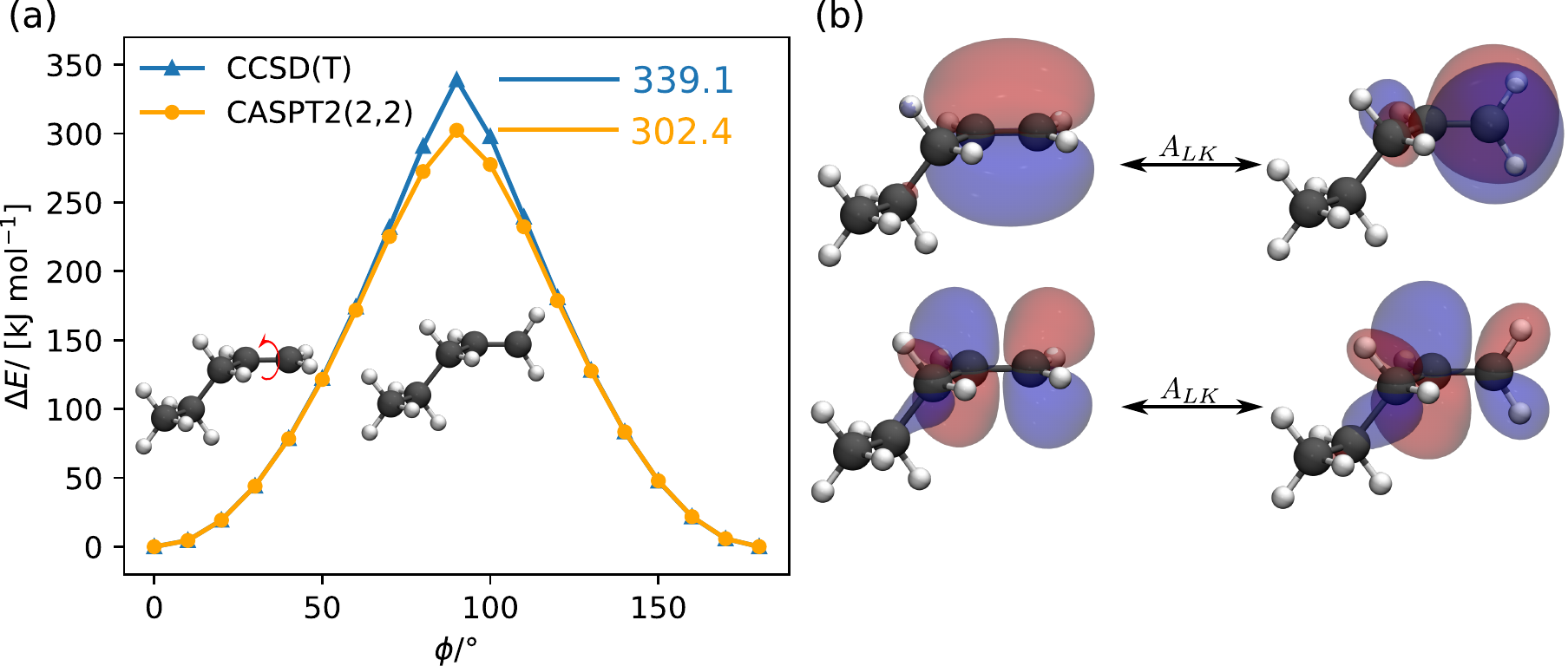}
    \caption{(a) Energy profile of the CH$_2$ group rotation around the carbon--carbon double bond of 1-pentene. The CAS(2,2) for CASPT2 was constructed by automatically selecting an active orbital space with \textsc{autoCAS} for the rotation angles $\phi = 0\si{\degree}$ and $\phi=90\si{\degree}$ and transferring the combined active space to all other structures. (b) Orbital map of the orbitals selected for the active orbital space by \textsc{autoCAS} and CH$_2$ rotation around the carbon--carbon double bond. Orbital isosurfaces are shown for a value of $\pm 0.025~$a.u.}
    \label{fig:double_bond_rotation}
\end{figure}

To illustrate our approach for a second reaction coordinate, we investigated the rotation of the CH$_2$ group around the carbon--carbon double bond in 1-pentene by varying the H-C-C-H dihedral angle $\phi$ between $0\si{\degree}$ and $180\si{\degree}$, as illustrated in Fig.~\ref{fig:double_bond_rotation}(a). The active space selection was performed only for the angles $\phi=0\si{\degree}$ and $90\si{\degree}$ from the aligned and localized HF orbitals calculated in the cc-pVDZ basis set\cite{Dunning1989}. The combined active orbital spaces were then transferred to all other structures through the orbital mapping procedure (occupied: $\tau_\mathrm{min} = 0.5$, virtual: $\tau_\mathrm{min} = 0.5$). The bonding and anti-bonding $\pi$-orbitals were selected by \textsc{autoCAS} for the active orbital space of both structures. For these orbitals no match could be realized during the mapping procedure, leading to their assignment to the orbital sets $\{\psi_{iL}\}^\text{no-match}$. The isosurface plots of the $\pi$-orbitals are shown in Fig.~\ref{fig:double_bond_rotation}(b). The energy profile of the rotation of the CH$_2$ group is shown in Fig.~\ref{fig:double_bond_rotation}, calculated with CASPT2 for this automatically selected and transferred active orbital space. 

Furthermore, we show the energy profile for this rotation calculated with CCSD(T).
CCSD(T) predicts the barrier height for the rotation to be $36.7~\si{kJ.mol^{-1}}$ higher than CASPT2. The point of the maximum energy corresponds to the dihedral angle of $\phi = 90\si{\degree}$. In this case, the electronic structure shows a significant multi-reference character, as illustrated by the high value of $0.1930$ for the $D1$ diagnostic\cite{Janssen1998} calculated from the CCSD amplitudes (values of $D1 < 0.05$ indicate a single reference character of the wavefunction\cite{Janssen1998}), indicating that CCSD(T) is not a suitable wavefunction model to describe the electronic structure correctly for $\phi=90\si{\degree}$ and that CASPT2 should provide a more reliable description.
For angles from $0\si{\degree}$-$70\si{\degree}$ and $110\si{\degree}$-$180\si{\degree}$, the CASPT2 and CCSD(T) energy profiles agree well. Furthermore, the CASPT2 energy profile is smooth, showing that transferring active orbital spaces between structures with our mapping approach is reliable.

In this work, we demonstrated how occupied, and virtual orbitals can be matched between structures along a path on the Born--Oppenheimer hypersurface (such as a reaction coordinate). Our approach provides a rigorous protocol for constructing consistent active orbital spaces for CAS calculations for chemical reactions. The protocol requires no input arguments, is fully automatized, and is integrated with the automatic active orbital selection approach implemented in \textsc{autoCAS}. 

In future work, we will investigate the effect of tailoring the active space selection directly to the reaction by choosing the comparison threshold $\tau$ explicitly similar to the original embedding region selection presented in Ref.~\citenum{Bensberg2019a}. Moreover, we will consider the straightforward extension to electronically excited states.



\section*{Acknowledgment}
We thank Maximilian Mörchen for helpful discussions concerning the implementation of the interface between \textsc{autoCAS} and \textsc{Serenity}.

\providecommand{\latin}[1]{#1}
\makeatletter
\providecommand{\doi}
  {\begingroup\let\do\@makeother\dospecials
  \catcode`\{=1 \catcode`\}=2 \doi@aux}
\providecommand{\doi@aux}[1]{\endgroup\texttt{#1}}
\makeatother
\providecommand*\mcitethebibliography{\thebibliography}
\csname @ifundefined\endcsname{endmcitethebibliography}
  {\let\endmcitethebibliography\endthebibliography}{}


\begin{mcitethebibliography}{57}
\providecommand*\natexlab[1]{#1}
\providecommand*\mciteSetBstSublistMode[1]{}
\providecommand*\mciteSetBstMaxWidthForm[2]{}
\providecommand*\mciteBstWouldAddEndPuncttrue
  {\def\EndOfBibitem{\unskip.}}
\providecommand*\mciteBstWouldAddEndPunctfalse
  {\let\EndOfBibitem\relax}
\providecommand*\mciteSetBstMidEndSepPunct[3]{}
\providecommand*\mciteSetBstSublistLabelBeginEnd[3]{}
\providecommand*\EndOfBibitem{}
\mciteSetBstSublistMode{f}
\mciteSetBstMaxWidthForm{subitem}{(\alph{mcitesubitemcount})}
\mciteSetBstSublistLabelBeginEnd
  {\mcitemaxwidthsubitemform\space}
  {\relax}
  {\relax}

\bibitem[Roos \latin{et~al.}(1980)Roos, Taylor, and Siegbahn]{Roos1980}
Roos,~B.~O.; Taylor,~P.~R.; Siegbahn,~P.~E. A complete active space {SCF}
  method ({CASSCF}) using a density matrix formulated super-{CI} approach.
  \emph{Chem. Phys.} \textbf{1980}, \emph{48}, 157--173\relax
\mciteBstWouldAddEndPuncttrue
\mciteSetBstMidEndSepPunct{\mcitedefaultmidpunct}
{\mcitedefaultendpunct}{\mcitedefaultseppunct}\relax
\EndOfBibitem
\bibitem[Ruedenberg \latin{et~al.}(1982)Ruedenberg, Schmidt, Gilbert, and
  Elbert]{Ruedenberg1982}
Ruedenberg,~K.; Schmidt,~M.~W.; Gilbert,~M.~M.; Elbert,~S. Are atoms intrinsic
  to molecular electronic wavefunctions? I. The {FORS} model. \emph{Chem Phys}
  \textbf{1982}, \emph{71}, 41--49\relax
\mciteBstWouldAddEndPuncttrue
\mciteSetBstMidEndSepPunct{\mcitedefaultmidpunct}
{\mcitedefaultendpunct}{\mcitedefaultseppunct}\relax
\EndOfBibitem
\bibitem[Knowles and Werner(1985)Knowles, and Werner]{Knowles1985}
Knowles,~P.~J.; Werner,~H.-J. An efficient second-order {MC} {SCF} method for
  long configuration expansions. \emph{Chem. Phys. Lett.} \textbf{1985},
  \emph{115}, 259--267\relax
\mciteBstWouldAddEndPuncttrue
\mciteSetBstMidEndSepPunct{\mcitedefaultmidpunct}
{\mcitedefaultendpunct}{\mcitedefaultseppunct}\relax
\EndOfBibitem
\bibitem[Shepard(1987)]{Shepard1987}
Shepard,~R. The multiconfiguration self-consistent field method. \emph{Adv.
  Chem. Phys.} \textbf{1987}, \emph{69}, 63\relax
\mciteBstWouldAddEndPuncttrue
\mciteSetBstMidEndSepPunct{\mcitedefaultmidpunct}
{\mcitedefaultendpunct}{\mcitedefaultseppunct}\relax
\EndOfBibitem
\bibitem[Booth \latin{et~al.}(2009)Booth, Thom, and Alavi]{Booth2009}
Booth,~G.~H.; Thom,~A. J.~W.; Alavi,~A. Fermion Monte Carlo without fixed
  nodes: A game of life, death, and annihilation in Slater determinant space.
  \emph{J. Chem. Phys.} \textbf{2009}, \emph{131}, 054106\relax
\mciteBstWouldAddEndPuncttrue
\mciteSetBstMidEndSepPunct{\mcitedefaultmidpunct}
{\mcitedefaultendpunct}{\mcitedefaultseppunct}\relax
\EndOfBibitem
\bibitem[Booth and Alavi(2010)Booth, and Alavi]{Booth2010}
Booth,~G.~H.; Alavi,~A. Approaching chemical accuracy using full
  configuration-interaction quantum Monte Carlo: A study of ionization
  potentials. \emph{J. Chem. Phys.} \textbf{2010}, \emph{132}, 174104\relax
\mciteBstWouldAddEndPuncttrue
\mciteSetBstMidEndSepPunct{\mcitedefaultmidpunct}
{\mcitedefaultendpunct}{\mcitedefaultseppunct}\relax
\EndOfBibitem
\bibitem[White(1992)]{White1992}
White,~S.~R. Density matrix formulation for quantum renormalization groups.
  \emph{Phys. Rev. Lett.} \textbf{1992}, \emph{69}, 2863--2866\relax
\mciteBstWouldAddEndPuncttrue
\mciteSetBstMidEndSepPunct{\mcitedefaultmidpunct}
{\mcitedefaultendpunct}{\mcitedefaultseppunct}\relax
\EndOfBibitem
\bibitem[Baiardi and Reiher(2020)Baiardi, and Reiher]{Baiardi2020}
Baiardi,~A.; Reiher,~M. The density matrix renormalization group in chemistry
  and molecular physics: Recent developments and new challenges. \emph{J. Chem.
  Phys.} \textbf{2020}, \emph{152}, 040903\relax
\mciteBstWouldAddEndPuncttrue
\mciteSetBstMidEndSepPunct{\mcitedefaultmidpunct}
{\mcitedefaultendpunct}{\mcitedefaultseppunct}\relax
\EndOfBibitem
\bibitem[Lindh and Galv{\'{a}}n(2020)Lindh, and Galv{\'{a}}n]{Lindh2020}
Lindh,~R.; Galv{\'{a}}n,~I.~F. In \emph{Quantum Chemistry and Dynamics of
  Excited States}; Gonz{\'a}lez,~L., Lindh,~R., Eds.; Wiley, 2020; pp
  299--353\relax
\mciteBstWouldAddEndPuncttrue
\mciteSetBstMidEndSepPunct{\mcitedefaultmidpunct}
{\mcitedefaultendpunct}{\mcitedefaultseppunct}\relax
\EndOfBibitem
\bibitem[Stein and Reiher(2017)Stein, and Reiher]{Stein2017}
Stein,~C.~J.; Reiher,~M. Automated Identification of Relevant Frontier Orbitals
  for Chemical Compounds and Processes. \emph{{CHIMIA}} \textbf{2017},
  \emph{71}, 170--176\relax
\mciteBstWouldAddEndPuncttrue
\mciteSetBstMidEndSepPunct{\mcitedefaultmidpunct}
{\mcitedefaultendpunct}{\mcitedefaultseppunct}\relax
\EndOfBibitem
\bibitem[Pulay and Hamilton(1988)Pulay, and Hamilton]{Pulay1988}
Pulay,~P.; Hamilton,~T.~P. {UHF} natural orbitals for defining and starting
  {MC}-{SCF} calculations. \emph{J. Chem. Phys.} \textbf{1988}, \emph{88},
  4926--4933\relax
\mciteBstWouldAddEndPuncttrue
\mciteSetBstMidEndSepPunct{\mcitedefaultmidpunct}
{\mcitedefaultendpunct}{\mcitedefaultseppunct}\relax
\EndOfBibitem
\bibitem[Bofill and Pulay(1989)Bofill, and Pulay]{Bofill1989}
Bofill,~J.~M.; Pulay,~P. The unrestricted natural orbital{\textendash}complete
  active space ({UNO}{\textendash}{CAS}) method: An inexpensive alternative to
  the complete active space{\textendash}self-consistent-field
  ({CAS}{\textendash}{SCF}) method. \emph{J. Chem. Phys.} \textbf{1989},
  \emph{90}, 3637--3646\relax
\mciteBstWouldAddEndPuncttrue
\mciteSetBstMidEndSepPunct{\mcitedefaultmidpunct}
{\mcitedefaultendpunct}{\mcitedefaultseppunct}\relax
\EndOfBibitem
\bibitem[Keller \latin{et~al.}(2015)Keller, Boguslawski, Janowski, Reiher, and
  Pulay]{Keller2015a}
Keller,~S.; Boguslawski,~K.; Janowski,~T.; Reiher,~M.; Pulay,~P. Selection of
  active spaces for multiconfigurational wavefunctions. \emph{J. Chem. Phys.}
  \textbf{2015}, \emph{142}, 244104\relax
\mciteBstWouldAddEndPuncttrue
\mciteSetBstMidEndSepPunct{\mcitedefaultmidpunct}
{\mcitedefaultendpunct}{\mcitedefaultseppunct}\relax
\EndOfBibitem
\bibitem[Jensen \latin{et~al.}(1988)Jensen, J{\o}rgensen, {\AA}gren, and
  Olsen]{Jensen1988}
Jensen,~H. J.~A.; J{\o}rgensen,~P.; {\AA}gren,~H.; Olsen,~J. {Second-order
  M{\o}ller--Plesset perturbation theory as a configuration and orbital
  generator in multiconfiguration self-consistent field calculations}. \emph{J.
  Chem. Phys.} \textbf{1988}, \emph{88}, 3834--3839\relax
\mciteBstWouldAddEndPuncttrue
\mciteSetBstMidEndSepPunct{\mcitedefaultmidpunct}
{\mcitedefaultendpunct}{\mcitedefaultseppunct}\relax
\EndOfBibitem
\bibitem[Stein and Reiher(2016)Stein, and Reiher]{Stein2016}
Stein,~C.~J.; Reiher,~M. Automated Selection of Active Orbital Spaces. \emph{J.
  Chem. Theory Comput.} \textbf{2016}, \emph{12}, 1760--1771\relax
\mciteBstWouldAddEndPuncttrue
\mciteSetBstMidEndSepPunct{\mcitedefaultmidpunct}
{\mcitedefaultendpunct}{\mcitedefaultseppunct}\relax
\EndOfBibitem
\bibitem[Legeza and S{\'{o}}lyom(2003)Legeza, and S{\'{o}}lyom]{Legeza2003}
Legeza,~O.; S{\'{o}}lyom,~J. Optimizing the density-matrix renormalization
  group method using quantum information entropy. \emph{Phys. Rev. B}
  \textbf{2003}, \emph{68}, 195116\relax
\mciteBstWouldAddEndPuncttrue
\mciteSetBstMidEndSepPunct{\mcitedefaultmidpunct}
{\mcitedefaultendpunct}{\mcitedefaultseppunct}\relax
\EndOfBibitem
\bibitem[Stein \latin{et~al.}(2016)Stein, von Burg, and Reiher]{Stein2016a}
Stein,~C.~J.; von Burg,~V.; Reiher,~M. The Delicate Balance of Static and
  Dynamic Electron Correlation. \emph{J. Chem. Theory Comput.} \textbf{2016},
  \emph{12}, 3764--3773\relax
\mciteBstWouldAddEndPuncttrue
\mciteSetBstMidEndSepPunct{\mcitedefaultmidpunct}
{\mcitedefaultendpunct}{\mcitedefaultseppunct}\relax
\EndOfBibitem
\bibitem[Stein and Reiher(2019)Stein, and Reiher]{Stein2019}
Stein,~C.~J.; Reiher,~M. \textsc{autoCAS}: A Program for Fully Automated
  Multiconfigurational Calculations. \emph{J. Comput. Chem.} \textbf{2019},
  \relax
\mciteBstWouldAddEndPunctfalse
\mciteSetBstMidEndSepPunct{\mcitedefaultmidpunct}
{}{\mcitedefaultseppunct}\relax
\EndOfBibitem
\bibitem[Unsleber \latin{et~al.}(2022)Unsleber, Liu, Talirz, Weymuth, Mörchen,
  Grofe, Wecker, Stein, Panyala, Peng, Kowalski, Troyer, and
  Reiher]{Unsleber2022a}
Unsleber,~J.~P.; Liu,~H.; Talirz,~L.; Weymuth,~T.; Mörchen,~M.; Grofe,~A.;
  Wecker,~D.; Stein,~C.~J.; Panyala,~A.; Peng,~B.; Kowalski,~K.; Troyer,~M.;
  Reiher,~M. High-throughput ab initio reaction mechanism exploration in the
  cloud with automated multi-reference validation. 2022; DOI:
  10.48550/ARXIV.2211.14688\relax
\mciteBstWouldAddEndPuncttrue
\mciteSetBstMidEndSepPunct{\mcitedefaultmidpunct}
{\mcitedefaultendpunct}{\mcitedefaultseppunct}\relax
\EndOfBibitem
\bibitem[Sayfutyarova \latin{et~al.}(2017)Sayfutyarova, Sun, Chan, and
  Knizia]{Sayfutyarova2017}
Sayfutyarova,~E.~R.; Sun,~Q.; Chan,~G. K.-L.; Knizia,~G. Automated Construction
  of Molecular Active Spaces from Atomic Valence Orbitals. \emph{J. Chem.
  Theory Comput.} \textbf{2017}, \emph{13}, 4063--4078\relax
\mciteBstWouldAddEndPuncttrue
\mciteSetBstMidEndSepPunct{\mcitedefaultmidpunct}
{\mcitedefaultendpunct}{\mcitedefaultseppunct}\relax
\EndOfBibitem
\bibitem[Bao \latin{et~al.}(2018)Bao, Dong, Gagliardi, and Truhlar]{Bao2018}
Bao,~J.~J.; Dong,~S.~S.; Gagliardi,~L.; Truhlar,~D.~G. Automatic Selection of
  an Active Space for Calculating Electronic Excitation Spectra by
  {MS}-{CASPT}2 or {MC}-{PDFT}. \emph{J. Chem. Theory Comput.} \textbf{2018},
  \emph{14}, 2017--2025\relax
\mciteBstWouldAddEndPuncttrue
\mciteSetBstMidEndSepPunct{\mcitedefaultmidpunct}
{\mcitedefaultendpunct}{\mcitedefaultseppunct}\relax
\EndOfBibitem
\bibitem[Khedkar and Roemelt(2019)Khedkar, and Roemelt]{Khedkar2019}
Khedkar,~A.; Roemelt,~M. Active Space Selection Based on Natural Orbital
  Occupation Numbers from $n$-Electron Valence Perturbation Theory. \emph{J.
  Chem. Theory Comput.} \textbf{2019}, \emph{15}, 3522--3536\relax
\mciteBstWouldAddEndPuncttrue
\mciteSetBstMidEndSepPunct{\mcitedefaultmidpunct}
{\mcitedefaultendpunct}{\mcitedefaultseppunct}\relax
\EndOfBibitem
\bibitem[Welborn \latin{et~al.}(2018)Welborn, Manby, and Miller]{Welborn2018}
Welborn,~M.; Manby,~F.~R.; Miller,~T.~F. Even-handed subsystem selection in
  projection-based embedding. \emph{J. Chem. Phys.} \textbf{2018}, \emph{149},
  144101\relax
\mciteBstWouldAddEndPuncttrue
\mciteSetBstMidEndSepPunct{\mcitedefaultmidpunct}
{\mcitedefaultendpunct}{\mcitedefaultseppunct}\relax
\EndOfBibitem
\bibitem[Bensberg and Neugebauer(2019)Bensberg, and Neugebauer]{Bensberg2019a}
Bensberg,~M.; Neugebauer,~J. Direct orbital selection for projection-based
  embedding. \emph{J. Chem. Phys.} \textbf{2019}, \emph{150}, 214106\relax
\mciteBstWouldAddEndPuncttrue
\mciteSetBstMidEndSepPunct{\mcitedefaultmidpunct}
{\mcitedefaultendpunct}{\mcitedefaultseppunct}\relax
\EndOfBibitem
\bibitem[Manby \latin{et~al.}(2012)Manby, Stella, Goodpaster, and
  Miller]{Manby2012}
Manby,~F.~R.; Stella,~M.; Goodpaster,~J.~D.; Miller,~T.~F. A Simple, Exact
  Density-Functional-Theory Embedding Scheme. \emph{J. Chem. Theory Comput.}
  \textbf{2012}, \emph{8}, 2564--2568\relax
\mciteBstWouldAddEndPuncttrue
\mciteSetBstMidEndSepPunct{\mcitedefaultmidpunct}
{\mcitedefaultendpunct}{\mcitedefaultseppunct}\relax
\EndOfBibitem
\bibitem[Sparta \latin{et~al.}(2017)Sparta, Retegan, Pinski, Riplinger, Becker,
  and Neese]{Sparta2017}
Sparta,~M.; Retegan,~M.; Pinski,~P.; Riplinger,~C.; Becker,~U.; Neese,~F.
  Multilevel Approaches within the Local Pair Natural Orbital Framework.
  \emph{J. Chem. Theory Comput.} \textbf{2017}, \emph{13}, 3198--3207\relax
\mciteBstWouldAddEndPuncttrue
\mciteSetBstMidEndSepPunct{\mcitedefaultmidpunct}
{\mcitedefaultendpunct}{\mcitedefaultseppunct}\relax
\EndOfBibitem
\bibitem[Mata \latin{et~al.}(2008)Mata, Werner, and Sch{\"u}tz]{Mata2008}
Mata,~R.~A.; Werner,~H.-J.; Sch{\"u}tz,~M. Correlation regions within a
  localized molecular orbital approach. \emph{J. Chem. Phys.} \textbf{2008},
  \emph{128}, 144106\relax
\mciteBstWouldAddEndPuncttrue
\mciteSetBstMidEndSepPunct{\mcitedefaultmidpunct}
{\mcitedefaultendpunct}{\mcitedefaultseppunct}\relax
\EndOfBibitem
\bibitem[Li and Piecuch(2010)Li, and Piecuch]{Li2010a}
Li,~W.; Piecuch,~P. Multilevel Extension of the Cluster-in-Molecule Local
  Correlation Methodology: Merging Coupled-Cluster and M{\o}ller-Plesset
  Perturbation Theories. \emph{J. Phys. Chem. A} \textbf{2010}, \emph{114},
  6721--6727\relax
\mciteBstWouldAddEndPuncttrue
\mciteSetBstMidEndSepPunct{\mcitedefaultmidpunct}
{\mcitedefaultendpunct}{\mcitedefaultseppunct}\relax
\EndOfBibitem
\bibitem[Rolik and K{\'{a}}llay(2011)Rolik, and K{\'{a}}llay]{Rolik2011}
Rolik,~Z.; K{\'{a}}llay,~M. A general-order local coupled-cluster method based
  on the cluster-in-molecule approach. \emph{J. Chem. Phys.} \textbf{2011},
  \emph{135}, 104111\relax
\mciteBstWouldAddEndPuncttrue
\mciteSetBstMidEndSepPunct{\mcitedefaultmidpunct}
{\mcitedefaultendpunct}{\mcitedefaultseppunct}\relax
\EndOfBibitem
\bibitem[Barnes \latin{et~al.}(2019)Barnes, Bykov, Lyakh, and
  Straatsma]{Barnes2019}
Barnes,~A.~L.; Bykov,~D.; Lyakh,~D.~I.; Straatsma,~T.~P. Multilayer
  Divide-Expand-Consolidate Coupled-Cluster Method: Demonstrative Calculations
  of the Adsorption Energy of Carbon Dioxide in the Mg-{MOF}-74
  Metal{\textendash}Organic Framework. \emph{J. Phys. Chem. A} \textbf{2019},
  \emph{123}, 8734--8743\relax
\mciteBstWouldAddEndPuncttrue
\mciteSetBstMidEndSepPunct{\mcitedefaultmidpunct}
{\mcitedefaultendpunct}{\mcitedefaultseppunct}\relax
\EndOfBibitem
\bibitem[Bensberg and Neugebauer(2020)Bensberg, and Neugebauer]{Bensberg2020a}
Bensberg,~M.; Neugebauer,~J. Density functional theory based embedding
  approaches for transition-metal complexes. \emph{Phys. Chem. Chem. Phys.}
  \textbf{2020}, \emph{22}, 26093--26103\relax
\mciteBstWouldAddEndPuncttrue
\mciteSetBstMidEndSepPunct{\mcitedefaultmidpunct}
{\mcitedefaultendpunct}{\mcitedefaultseppunct}\relax
\EndOfBibitem
\bibitem[Bensberg and Neugebauer(2021)Bensberg, and Neugebauer]{Bensberg2021a}
Bensberg,~M.; Neugebauer,~J. Direct orbital selection within the domain-based
  local pair natural orbital coupled-cluster method. \emph{J. Chem. Phys.}
  \textbf{2021}, \emph{155}, 224102\relax
\mciteBstWouldAddEndPuncttrue
\mciteSetBstMidEndSepPunct{\mcitedefaultmidpunct}
{\mcitedefaultendpunct}{\mcitedefaultseppunct}\relax
\EndOfBibitem
\bibitem[Bensberg \latin{et~al.}(2022)Bensberg, Grimmel, Simm, Sobez, Steiner,
  Türtscher, Unsleber, Weymuth, and Reiher]{Bensberg2022b}
Bensberg,~M.; Grimmel,~S.~A.; Simm,~G.~N.; Sobez,~J.-G.; Steiner,~M.;
  Türtscher,~P.~L.; Unsleber,~J.~P.; Weymuth,~T.; Reiher,~M. qcscine/chemoton:
  Release 2.1.0. 2022; DOI: 10.5281/zenodo.6984579\relax
\mciteBstWouldAddEndPuncttrue
\mciteSetBstMidEndSepPunct{\mcitedefaultmidpunct}
{\mcitedefaultendpunct}{\mcitedefaultseppunct}\relax
\EndOfBibitem
\bibitem[Mörchen \latin{et~al.}(2022)Mörchen, Stein, Unsleber, and
  Reiher]{Moerchen2022}
Mörchen,~M.; Stein,~C.~J.; Unsleber,~J.~P.; Reiher,~M. qcscine/autocas:
  Release 2.0.0. 2022; DOI: 10.5281/ZENODO.7179860\relax
\mciteBstWouldAddEndPuncttrue
\mciteSetBstMidEndSepPunct{\mcitedefaultmidpunct}
{\mcitedefaultendpunct}{\mcitedefaultseppunct}\relax
\EndOfBibitem
\bibitem[Mulliken(1955)]{Mulliken1955}
Mulliken,~R.~S. Electronic Population Analysis on {LCAO}{\textendash}{MO}
  Molecular Wave Functions. I. \emph{J. Chem. Phys.} \textbf{1955}, \emph{23},
  1833--1840\relax
\mciteBstWouldAddEndPuncttrue
\mciteSetBstMidEndSepPunct{\mcitedefaultmidpunct}
{\mcitedefaultendpunct}{\mcitedefaultseppunct}\relax
\EndOfBibitem
\bibitem[Knizia(2013)]{Knizia2013}
Knizia,~G. Intrinsic Atomic Orbitals: An Unbiased Bridge between Quantum Theory
  and Chemical Concepts. \emph{J. Chem. Theory Comput.} \textbf{2013},
  \emph{9}, 4834--4843\relax
\mciteBstWouldAddEndPuncttrue
\mciteSetBstMidEndSepPunct{\mcitedefaultmidpunct}
{\mcitedefaultendpunct}{\mcitedefaultseppunct}\relax
\EndOfBibitem
\bibitem[Senjean \latin{et~al.}(2021)Senjean, Sen, Repisky, Knizia, and
  Visscher]{Senjean2021}
Senjean,~B.; Sen,~S.; Repisky,~M.; Knizia,~G.; Visscher,~L. Generalization of
  Intrinsic Orbitals to Kramers-Paired Quaternion Spinors, Molecular Fragments,
  and Valence Virtual Spinors. \emph{J Chem. Theory Comput.} \textbf{2021},
  \emph{17}, 1337--1354\relax
\mciteBstWouldAddEndPuncttrue
\mciteSetBstMidEndSepPunct{\mcitedefaultmidpunct}
{\mcitedefaultendpunct}{\mcitedefaultseppunct}\relax
\EndOfBibitem
\bibitem[Bensberg and Neugebauer(2020)Bensberg, and Neugebauer]{Bensberg2020}
Bensberg,~M.; Neugebauer,~J. Orbital Alignment for Accurate Projection-Based
  Embedding Calculations along Reaction Paths. \emph{J. Chem. Theory Comput.}
  \textbf{2020}, \emph{16}, 3607--3619\relax
\mciteBstWouldAddEndPuncttrue
\mciteSetBstMidEndSepPunct{\mcitedefaultmidpunct}
{\mcitedefaultendpunct}{\mcitedefaultseppunct}\relax
\EndOfBibitem
\bibitem[Keller \latin{et~al.}(2015)Keller, Dolfi, Troyer, and
  Reiher]{Keller2015}
Keller,~S.; Dolfi,~M.; Troyer,~M.; Reiher,~M. An efficient matrix product
  operator representation of the quantum chemical Hamiltonian. \emph{J. Chem.
  Phys.} \textbf{2015}, \emph{143}, 244118\relax
\mciteBstWouldAddEndPuncttrue
\mciteSetBstMidEndSepPunct{\mcitedefaultmidpunct}
{\mcitedefaultendpunct}{\mcitedefaultseppunct}\relax
\EndOfBibitem
\bibitem[Unsleber \latin{et~al.}(2018)Unsleber, Dresselhaus, Klahr, Schnieders,
  B{\"{o}}ckers, Barton, and Neugebauer]{Serenity2018}
Unsleber,~J.~P.; Dresselhaus,~T.; Klahr,~K.; Schnieders,~D.; B{\"{o}}ckers,~M.;
  Barton,~D.; Neugebauer,~J. Serenity : A subsystem quantum chemistry program.
  \emph{J. Comput. Chem.} \textbf{2018}, \emph{39}, 788--798\relax
\mciteBstWouldAddEndPuncttrue
\mciteSetBstMidEndSepPunct{\mcitedefaultmidpunct}
{\mcitedefaultendpunct}{\mcitedefaultseppunct}\relax
\EndOfBibitem
\bibitem[Niemeyer \latin{et~al.}(2022)Niemeyer, Eschenbach, Bensberg, Tölle,
  Hellmann, Lampe, Massolle, Rikus, Schnieders, Unsleber, and
  Neugebauer]{Niemeyer2022}
Niemeyer,~N.; Eschenbach,~P.; Bensberg,~M.; Tölle,~J.; Hellmann,~L.;
  Lampe,~L.; Massolle,~A.; Rikus,~A.; Schnieders,~D.; Unsleber,~J.~P.;
  Neugebauer,~J. The subsystem quantum chemistry program \textsc{Serenity}.
  \emph{WIREs Comput. Mol. Sci.} \textbf{2022}, e1647\relax
\mciteBstWouldAddEndPuncttrue
\mciteSetBstMidEndSepPunct{\mcitedefaultmidpunct}
{\mcitedefaultendpunct}{\mcitedefaultseppunct}\relax
\EndOfBibitem
\bibitem[Ser()]{SerenityGitHub}
Latest Release of \textsc{Serenity} is available from
  https://github.com/qcserenity/serenity\relax
\mciteBstWouldAddEndPuncttrue
\mciteSetBstMidEndSepPunct{\mcitedefaultmidpunct}
{\mcitedefaultendpunct}{\mcitedefaultseppunct}\relax
\EndOfBibitem
\bibitem[Vaucher and Reiher(2018)Vaucher, and Reiher]{Vaucher2018}
Vaucher,~A.~C.; Reiher,~M. Minimum Energy Paths and Transition States by Curve
  Optimization. \emph{J. Chem. Theory Comput.} \textbf{2018}, \emph{14},
  3091--3099\relax
\mciteBstWouldAddEndPuncttrue
\mciteSetBstMidEndSepPunct{\mcitedefaultmidpunct}
{\mcitedefaultendpunct}{\mcitedefaultseppunct}\relax
\EndOfBibitem
\bibitem[Brunken \latin{et~al.}(2022)Brunken, Csizi, Grimmel, Gugler, Sobez,
  Steiner, Türtscher, Unsleber, Vaucher, Weymuth, and Reiher]{Brunken2022}
Brunken,~C.; Csizi,~K.-S.; Grimmel,~S.~A.; Gugler,~S.; Sobez,~J.-G.;
  Steiner,~M.; Türtscher,~P.~L.; Unsleber,~J.~P.; Vaucher,~A.~C.; Weymuth,~T.;
  Reiher,~M. qcscine/readuct: Release 4.1.0. 2022; DOI:
  10.5281/zenodo.6984575\relax
\mciteBstWouldAddEndPuncttrue
\mciteSetBstMidEndSepPunct{\mcitedefaultmidpunct}
{\mcitedefaultendpunct}{\mcitedefaultseppunct}\relax
\EndOfBibitem
\bibitem[Perdew \latin{et~al.}(1996)Perdew, Burke, and Ernzerhof]{Perdew96}
Perdew,~J.~P.; Burke,~K.; Ernzerhof,~M. {Generalized Gradient Approximation
  Made Simple}. \emph{Phys. Rev. Lett.} \textbf{1996}, \emph{77}, 3865\relax
\mciteBstWouldAddEndPuncttrue
\mciteSetBstMidEndSepPunct{\mcitedefaultmidpunct}
{\mcitedefaultendpunct}{\mcitedefaultseppunct}\relax
\EndOfBibitem
\bibitem[Grimme \latin{et~al.}(2010)Grimme, Antony, Ehrlich, and
  Krieg]{Grimme2010a}
Grimme,~S.; Antony,~J.; Ehrlich,~S.; Krieg,~H. A consistent and accurate ab
  initio parametrization of density functional dispersion correction ({DFT}-D)
  for the 94 elements H-Pu. \emph{J. Chem. Phys.} \textbf{2010}, \emph{132},
  154104\relax
\mciteBstWouldAddEndPuncttrue
\mciteSetBstMidEndSepPunct{\mcitedefaultmidpunct}
{\mcitedefaultendpunct}{\mcitedefaultseppunct}\relax
\EndOfBibitem
\bibitem[Grimme \latin{et~al.}(2011)Grimme, Ehrlich, and Goerigk]{Grimme2011}
Grimme,~S.; Ehrlich,~S.; Goerigk,~L. {Effect of the damping function in
  dispersion corrected density functional theory}. \emph{J. Comput. Chem.}
  \textbf{2011}, \emph{32}, 1456--1465\relax
\mciteBstWouldAddEndPuncttrue
\mciteSetBstMidEndSepPunct{\mcitedefaultmidpunct}
{\mcitedefaultendpunct}{\mcitedefaultseppunct}\relax
\EndOfBibitem
\bibitem[Dunning(1989)]{Dunning1989}
Dunning,~T.~H. Gaussian basis sets for use in correlated molecular
  calculations. I. The atoms boron through neon and hydrogen. \emph{J. Chem.
  Phys.} \textbf{1989}, \emph{90}, 1007--1023\relax
\mciteBstWouldAddEndPuncttrue
\mciteSetBstMidEndSepPunct{\mcitedefaultmidpunct}
{\mcitedefaultendpunct}{\mcitedefaultseppunct}\relax
\EndOfBibitem
\bibitem[tur()]{turbomole741}
{TURBOMOLE V7.4.2 2019}, a development of {University of Karlsruhe} and
  {Forschungszentrum Karlsruhe GmbH}, 1989-2007, {TURBOMOLE GmbH}, since 2007;
  available from {\tt http://www.turbomole.com}.\relax
\mciteBstWouldAddEndPunctfalse
\mciteSetBstMidEndSepPunct{\mcitedefaultmidpunct}
{}{\mcitedefaultseppunct}\relax
\EndOfBibitem
\bibitem[Andersson \latin{et~al.}(1990)Andersson, Malmqvist, Roos, Sadlej, and
  Wolinski]{Andersson1990}
Andersson,~K.; Malmqvist,~P.~A.; Roos,~B.~O.; Sadlej,~A.~J.; Wolinski,~K.
  {Second-order perturbation theory with a CASSCF reference function}. \emph{J.
  Phys. Chem.} \textbf{1990}, \emph{94}, 5483--5488\relax
\mciteBstWouldAddEndPuncttrue
\mciteSetBstMidEndSepPunct{\mcitedefaultmidpunct}
{\mcitedefaultendpunct}{\mcitedefaultseppunct}\relax
\EndOfBibitem
\bibitem[Andersson \latin{et~al.}(1992)Andersson, Malmqvist, and
  Roos]{Andersson1992}
Andersson,~K.; Malmqvist,~P.-{\AA}.; Roos,~B.~O. Second-order perturbation
  theory with a complete active space self-consistent field reference function.
  \emph{J. Chem. Phys.} \textbf{1992}, \emph{96}, 1218--1226\relax
\mciteBstWouldAddEndPuncttrue
\mciteSetBstMidEndSepPunct{\mcitedefaultmidpunct}
{\mcitedefaultendpunct}{\mcitedefaultseppunct}\relax
\EndOfBibitem
\bibitem[Ghigo \latin{et~al.}(2004)Ghigo, Roos, and Malmqvist]{Ghigo2004}
Ghigo,~G.; Roos,~B.~O.; Malmqvist,~P.-{\AA}. {A modified definition of the
  zeroth-order Hamiltonian in multiconfigurational perturbation theory
  (CASPT2)}. \emph{Chem. Phys. Lett.} \textbf{2004}, \emph{396}, 142--149\relax
\mciteBstWouldAddEndPuncttrue
\mciteSetBstMidEndSepPunct{\mcitedefaultmidpunct}
{\mcitedefaultendpunct}{\mcitedefaultseppunct}\relax
\EndOfBibitem
\bibitem[Freitag \latin{et~al.}(2017)Freitag, Knecht, Angeli, and
  Reiher]{Freitag2017}
Freitag,~L.; Knecht,~S.; Angeli,~C.; Reiher,~M. {Multireference Perturbation
  Theory with Cholesky Decomposition for the Density Matrix Renormalization
  Group}. \emph{J. Chem. Theory Comput.} \textbf{2017}, \emph{13},
  451--459\relax
\mciteBstWouldAddEndPuncttrue
\mciteSetBstMidEndSepPunct{\mcitedefaultmidpunct}
{\mcitedefaultendpunct}{\mcitedefaultseppunct}\relax
\EndOfBibitem
\bibitem[Ma \latin{et~al.}(2017)Ma, Knecht, Keller, and Reiher]{Ma2017}
Ma,~Y.; Knecht,~S.; Keller,~S.; Reiher,~M. {Second-Order Self-Consistent-Field
  Density-Matrix Renormalization Group}. \emph{J. Chem. Theory Comput.}
  \textbf{2017}, \emph{13}, 2533--2549\relax
\mciteBstWouldAddEndPuncttrue
\mciteSetBstMidEndSepPunct{\mcitedefaultmidpunct}
{\mcitedefaultendpunct}{\mcitedefaultseppunct}\relax
\EndOfBibitem
\bibitem[Raghavachari \latin{et~al.}(1989)Raghavachari, Trucks, Pople, and
  Head-Gordon]{Raghavachari1989}
Raghavachari,~K.; Trucks,~G.~W.; Pople,~J.~A.; Head-Gordon,~M. A fifth-order
  perturbation comparison of electron correlation theories. \emph{Chem. Phys.
  Lett.} \textbf{1989}, \emph{157}, 479--483\relax
\mciteBstWouldAddEndPuncttrue
\mciteSetBstMidEndSepPunct{\mcitedefaultmidpunct}
{\mcitedefaultendpunct}{\mcitedefaultseppunct}\relax
\EndOfBibitem
\bibitem[Janssen and Nielsen(1998)Janssen, and Nielsen]{Janssen1998}
Janssen,~C.~L.; Nielsen,~I.~M. New diagnostics for coupled-cluster and
  M{\o}ller{\textendash}Plesset perturbation theory. \emph{Chem. Phys. Lett.}
  \textbf{1998}, \emph{290}, 423--430\relax
\mciteBstWouldAddEndPuncttrue
\mciteSetBstMidEndSepPunct{\mcitedefaultmidpunct}
{\mcitedefaultendpunct}{\mcitedefaultseppunct}\relax
\EndOfBibitem
\end{mcitethebibliography}
\end{document}